\documentclass{article}

\usepackage{amssymb}
\usepackage{amsmath}
\usepackage{graphicx}
\usepackage{color}
\usepackage{soul}
\usepackage[american]{babel}
\usepackage{float}
\usepackage{listings}
\usepackage{subfloat}
\usepackage{algpseudocode}
\usepackage{algorithmicx}
\usepackage{algorithm}
\usepackage{url}

\newfloat{query}{thp}{lop}
\floatname{query}{Query}

\newfloat{table}{thp}{lop}
\floatname{table}{Table}

\definecolor{light-gray}{gray}{0.95}
\definecolor{light-blue}{cmyk}{0.1,0,0,0}
\newlength{\figurewidthA}
\begin{document}

\title{Efficient classification of billions of points into complex geographic regions using hierarchical triangular mesh}

\author{
	D\'aniel Kondor, L\'aszl\'o Dobos, Istv\'an Csabai, \\
	Andr\'as Bodor,	G\'abor Vattay \\
	\small{E\"otv\"os Lor\'and University, Department of Physics of Complex Systems}\\
	Tam\'as Budav\'ari and Alexander S. Szalay \\
	\small{The Johns Hopkins University, Department of Physics \& Astronomy}
}
%

\date{\today}


\maketitle

\begin{abstract}

We present a case study about the spatial indexing and regional classification of billions of geographic coordinates from geo-tagged social network data using Hierarchical Triangular Mesh (HTM) implemented for Microsoft SQL Server. Due to the lack of certain features of the HTM library, we use it in conjunction with the GIS functions of SQL Server to significantly increase the efficiency of pre-filtering of spatial filter and join queries. For example, we implemented a new algorithm to compute the HTM tessellation of complex geographic regions and precomputed the intersections of HTM triangles and geographic regions for faster false-positive filtering. With full control over the index structure, HTM-based pre-filtering of simple containment searches outperforms SQL Server spatial indices by a factor of ten and HTM-based spatial joins run about a hundred times faster.

\end{abstract}


\section{Introduction}

Spatial indexing of geographic data has always been an important component of database systems. Since the wide-spread adoption of social media and social networks, the size of data to be indexed has grown by multiple orders of magnitude, making even more demand on the efficiency of indexing algorithms and index structures. Certain fields of natural sciences also face similar problems: Astronomers, for example, have to perform spatial searches in databases of billions of observations where spatial criteria can be arbitrarily complex spherical regions. Inspired by astronomical problems, Szalay et al. \cite{htm, htm2} came up with a solution to index spatial data stored in Microsoft SQL Server years before native support of geographic indexing appeared in the product. Their indexing scheme is called Hierarchical Triangular Mesh (HTM) and uses an iteratively refined triangulation of the surface of the sphere to build a quad-tree index. In the present case study, we demonstrate how we applied HTM to index real GPS coordinates collected from the open data streams of Twitter, and performed a huge spatial join on the data to classify the coordinates by the administrative regions of the world. Our results show that pre-filtering capabilites of HTM are significantly better than that of the built-in spatial index of SQL Server and that HTM renders spatial joins originally thought impossible to be able to be computed in a reasonable time.  Source code and full queries are available at the following url: \url{http://www.vo.elte.hu/htmpaper}.

\section{Data and indexing}

Our data set consists of short messages (tweets) collected over a period of two years from the publicly available ``sprinkler'' data stream of Twitter. More than half of the tweets, over a billion, are geo-tagged. We built a Microsoft SQL Server database of the tweets \cite{twitterdb} and wanted to classify the geo-tagged messages by political administrative regions to investigate the geographic embedding of the social network of Twitter users.

Microsoft SQL Server supports spatial indexing since version~2008 via a library implemented on top of the integrated .Net framework runtime (SQL CLR). The library works by projecting the hemisphere onto a quadrilateral pyramid, then projecting the pyramid onto a square, and tessellating the square using four levels of fix-sized rectangular grids to construct a spatial index. The number of grid cells can be set between $4 \times 4$ and $16 \times 16$ providing a maximal index depth of 32~bits. The index structure itself is materialized as a hidden, but otherwise normal database table containing one row for each cell touched by the geography object being indexed. The table uses 11~bytes for the spatial index which is complemented by the primary key of the indexed object, in our case an additional 10~bytes. The final index size can be controlled by limiting the number of cells stored for each geography object resulting in less effective pre-filtering of spatial matches when the index size is kept small. The hard limit on the number of index entries per geography object is 8192.

Similarly to the built-in spatial index of SQL~Server, HTM indexing is also implemented in .Net. By default, HTM calculates a 40~bit deep hash, the so called HTM ID, from the coordinates. The 40~bit hash length corresponds to an almost uniform resolution of 10~meters on the surface of the Earth. HTM tessellates two dimensional shapes with small spherical triangles, called trixels. Trixels are represented by integer intervals; all coordinates with an HTM ID falling into the interval are guaranteed to be covered by the trixel. As HTM is a custom library, we had full control over the index structures, therefore we simply stored the 8~byte~HTM identifiers in the same table where the tweets were, and built an auxiliary index on the table ordered by the HTM identifier. Together with the primary key, the auxiliary index size was 18~bytes per row, exactly one row per coordinate pair.

We classified tweets using the maps from \url{gadm.org}, an open database of global administrative regions of every country in the world. The maps were loaded into the database as geography objects using the built-in geography type. For indexing, however, we decided to use HTM. This raised a problem because the HTM library was built for astronomical applications where regions on the sphere are better represented as unions of convex shapes contoured by great \textit{or} small circles than by vertices of polygons connected by great circles. This union-of-convexes representation is not appropriate for highly detailed complex maps as shapes would need to be decomposed into convexes first, a process that significantly increases the size of the data structures.

\begin{figure}
	\centering
	\includegraphics[width=\figurewidthA]{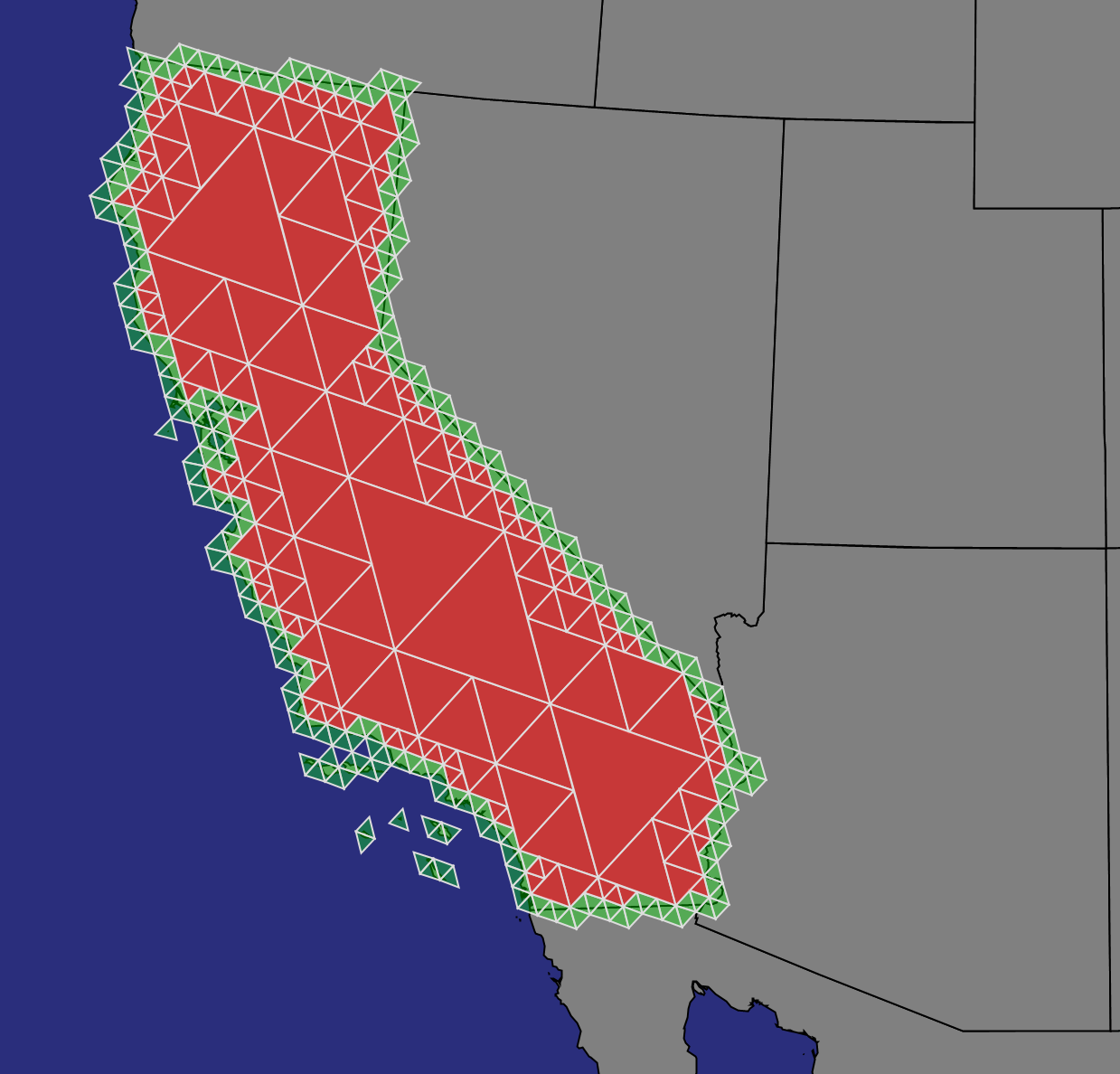}
	\caption{HTM cover of the U.S. state of California with level~9 trixels.}
	\label{fig:cali}
\end{figure}

Unfortunately, no code exists to directly compute the HTM tessellation of maps in the polygon representation, so we had to use another approach. By combining the HTM library and the built-in geographic library of SQL~Server, we determined the approximate HTM tessellation of regions up to a given precision by iteratively refining HTM trixels at the boundaries. Our solution, see Algorithm~\ref{lst:alg}, goes as follows. We construct a coarse tessellation of the region based on the bounding circle, then we intersect each trixel with the region using the built-in functions of SQL Server. If a trixel is completely inside the region it is added to the result set. Similarly, a completely disjoint trixel is discarded. Trixels intersecting with the boundary are refined into four sub-trixels and the algorithm is called recursively. Passing only the intersection of the original map with the trixel to the recursive call reduces the total runtime of the tessellation significantly. The algorithm uses the maximum depth of HTM trixels as a stop condition to limit the resolution of the tessellation but can be easily modified to use an upper limit on the number of trixels instead. Also, instead of trying to keep the index tables small, we store every trixel of the tessellation. Trixels on the deepest level which intersect with the boundary of the geography object are flagged as ``partial''. Figure~\ref{fig:cali} illustrates the results of the level~9 HTM covering of California with partial trixels in green. 

\begin{algorithm}
\begin{algorithmic}
	\Function{EvalTrixels}{region, trixellist, maxlevel}
		\State retlist $\leftarrow \varnothing$
		\ForAll{t \textbf{in} trixellist}
			\If{region.STContains(t)}				\Comment{Full trixel}
				\State t.Partial $\leftarrow$ false \Comment{Flag as full}
				\State retlist.Add(t)
			\Else \Comment{Partial or disjoint trixel}
				\State region2 $\leftarrow$ region.STIntersection(t)
				\If{region2 $\neq \varnothing$} 	\Comment{Partial trixel}
					\If{t.Level $\geq$ maxlevel}
						\State t.Partial $\leftarrow$ true \Comment{Flag as partial}
						\State retlist.Add(t) 
					\Else 
						\State tlist2 $\leftarrow$ t.Refine(t.Level+1) 
						\State retlist.AddRange(\Call{EvalTrixels}{region2, tlist2, maxlevel})
					\EndIf
				\EndIf
			\EndIf
		\EndFor
		\State \Return retlist
	\EndFunction
\end{algorithmic}
\caption{The function used for creating the HTM tessellation of a region. The function parameters are the region to be tessellated (region), the list of covering trixels to be refined (trixellist), and the maximum depth (maxlevel). The \texttt{STContains} and \texttt{STIntersection} functions are provided by the SQL~Server geography library. Trixels added to the result set are flagged as either full or partial.}
\label{lst:alg}
\end{algorithm}

\section{Spatial joins}

\begin{query}
\begin{lstlisting}
CREATE TABLE tweet (
    ID bigint PRIMARY KEY,
    HTMID bigint,
    coord geography )
    CREATE INDEX IX_tweet_htm
        ON tweet ( HTMID )
    CREATE SPATIAL INDEX IX_tweet_spatial
        ON tweet ( coord )
CREATE TABLE region (
    ID int PRIMARY KEY,
    geo geography )    
    CREATE SPATIAL INDEX IX_region_spatial
        ON region ( geo )
CREATE TABLE regionHTM (
    ID int,     --foreign key to region.ID
    start bigint,
    end bigint,
    partial bit )
\end{lstlisting}
\caption{Simplified schema of our database used for benchmarking HTM.}
\label{query:schema}
\end{query}

In order to explain the internals of HTM indexing, we create the schema of our database with Query~\ref{query:schema}. The HTM-based pre-filtering of a spatial join between a table containing GPS coordinates and another containing the tessellations of complex regions requires an inner join with a \texttt{BETWEEN} operator in the join constraint. Query~\ref{query:htmfilter} is a simplified example of such pre-filtering query. We will refer to these types of queries as \textit{range joins}. As range joins are highly optimized in the database engine, we expect excellent pre-filtering performance. The \texttt{LOOP JOIN} hint is added to suggest a query plan that consist of scan of the \texttt{tweet} table, while doing index seeks on the much smaller \texttt{regionHTM} table. This is optimal as long as the HTM index of the regions can be kept in memory. Using the built-in spatial index of SQL Server, pre-filtering of a spatial join can be done with Query~\ref{query:sqlfilter}. It translates into a rather complex execution plan that uses the spatial index for table \texttt{tweet} only, while calculating the tessellation of the geography objects in table \texttt{region} during query execution, or vice versa. By specifying query hints, one can tell the server which spatial index to use, but it seems impossible to use the spatial indices on \textit{both} tables at the same time. This behavior has a tremendous impact on the performance of spatial joins when using the built-in indices.

In case of the SQL Server geography index, exact containment testing can be done by simply replacing the function call to \texttt{Filter} with \texttt{STContains}, as in Query~\ref{query:sqlcontains}. When using the HTM index, points in full trixels are already accurately classified with Query~\ref{query:htmfilter}, only points in partial trixels need further processing to filter out false positive matches. This is done in Query~\ref{query:htmcontains} which again relies on the spatial functions of SQL Server. Also note, that Query~\ref{query:htmcontains}, by referencing the column \texttt{region.geo}, uses the entire region for testing point containment. In case of computing the spatial join of billions of coordinates with a limited number of complex regions, it is well worth to pre-compute the intersections of partial trixels and regions first and use them for containment testing instead of the whole regions. Due to publicational constraints we omit the query but all performance metrics quoted in the paper are measured using the pre-computed intersections of the regions and partial trixels for exact containment testing.

\section{Performance evaluation}

We measured the index generation time for the 50 continental states (plus Washington D.C.) of the United States using two different depths ($8 \times 8$ and $16 \times 16$ grids) of the SQL Server geography index and three different depths (level 12, 14 and 16) of HTM. For comparison, the $8 \times 8$ resolution of the SQL Server index roughly corresponds to a level 12 HTM index and the $16 \times 16$ grid resolution corresponds to a level 16 HTM index. The benchmarks were run on a 16-core database server with 96~GB memory. As our dataset fits into memory, queries are basically CPU-limited. Index generation times are summarized in Table~\ref{tab:idxgen}. Note, that SQL Server executed the geography index generation on two threads, while HTM ranges were generated on a single thread only. While we have no control over the internals of geography indices, the iterative refining of HTM tessellation could be replaced with a smarter, multi-threaded one. Also, the size of the geography indices is internally limited to 8192 entries per region, while the HTM indices were calculated without pruning, ultimately resulting in much larger index sizes.

\begin{query}
\begin{lstlisting}
SELECT t.ID, h.ID, h.partial
FROM tweet t
INNER LOOP JOIN regionHTM h
  ON t.htmID BETWEEN h.start AND h.end
\end{lstlisting}
\caption{Pre-filtering of a spatial join with HTM index.}
\label{query:htmfilter}
\end{query}

\begin{query}
\begin{lstlisting}
SELECT t.ID, r.ID
FROM tweet t INNER JOIN region r
  ON r.geo.Filter(t.coord) = 1
\end{lstlisting}
\label{query:sqlfilter}
\caption{Pre-filtering of a spatial join with the SQL Server geography index.}
\end{query}

\begin{query}
\begin{lstlisting}
SELECT tweet.ID, region.ID
FROM tweet t INNER JOIN region r
  ON r.geo.STContains(t.coord) = 1
\end{lstlisting}
\caption{Exact containment testing with the SQL Server geography index.}
\label{query:sqlcontains}
\end{query}

\begin{query}[h!]
\begin{lstlisting}
SELECT tweet.ID, regionHTM.ID
FROM tweet t INNER LOOP JOIN regionHTM h
  ON t.HTMID BETWEEN h.start AND h.end
INNER JOIN region r ON r.ID = h.ID
WHERE h.partial = 0 OR
      r.geo.STContains(t.coord) = 1
\end{lstlisting}
\caption{Classifying points from partial trixels.}
\label{query:htmcontains}
\end{query}

It is also rather instructive to compare the two indexing schemes by the false positive rates of pre-filtering. The results are listed in Table~\ref{tab:fp}. False positives rates of the HTM index are significantly lower in all cases, especially for higher index depths. In the case of the SQL Server spatial index, increasing the resolution does not help, but the opposite: it just makes things worse, as the number of index rows (and the resolution of the tessellation) is limited to a maximum of 8192 cells, insufficiently small for complex maps. Strictly limiting index size only helps when the number of shapes to be indexed is large and the shapes are relatively small and simple. When spatial indices fit into the memory, or at least can be read quickly from the disk, pre-filtering using range joins is expected to be significantly faster, even for indices with millions of rows, rather than exact containment testing against complex shapes. 

To test the performance of point classification, we prepared three samples having approximately 300~thousand, 1~million, 5~million points in each, uniformly sampled from the original database. Some tests were also run using the entire data set of more than one billion tweets. The coordinates covered the entire world but the majority of them were within the continental United States. The geographical distribution of the samples is realistic and follows the population density weighted by the local Twitter usage rate. To evaluate index performance, we computed a spatial join between a the samples of GPS coordinates with different cardinality and the 51 regions, first with pre-filtering only, then with exact containment test. Results of pre-filtering are listed in Table~\ref{tab:prefilter}, while exact containment metrics are shown in Table~\ref{tab:total}. All queries were executed using cold buffers, thus include I/O and CPU times.

The spatial join performance of the HTM index turned out to be about a hundred times better than the performance of the built-in geography index. Pre-filtering itself is about a thousand times faster than the built-in index, and usually could be done in a few seconds for the smaller samples. Such short query times are hard to be measured correctly, and values show a significant scatter when the queries are repeated. The two main reasons behind the significantly better performance of HTM are: 1) HTM-based pre-filtering could benefit from the spatial index on both tables, whereas SQL Server's geography library only used the index for one of the tables and calculated the tessellation for the other table on the fly. 2) The extensive pruning of index entries resulted in a very high rate of false positives in case of SQL~Server's geography index. Because of the pruning, increasing the index resolution could not actually increase the resolution of the tessellation in case of the rather complex maps. By using a significantly larger, but still manageable index table, and by intersecting the trixels of the tessellations with the regions to reduce the complexity of exact containment testing, HTM indexing could reduce the cost of spatial joins to a minimum. Based on these results, it is clear that running the point classification using only the built-in geography index of SQL Server index is not a viable solution for any task similar to ours, namely, when the number of points is in the billions range. 

\begin{table}[t]
\centering
\begin{tabular}{l|r|r}
	index type               & time [s] & index rows \\ \hline
	geography $8 \times 8$   &   13,352 &   412,055  \\
	geography $16 \times 16$ &    6,215 &   410,040  \\
	HTM level 12             &    4,366 &   267,763  \\
	HTM level 14             &    5,151 & 1,331,632  \\
	HTM level 16             &    9,952 & 6,354,932
\end{tabular}
\caption{Index generation time and number of index rows of the regions.}
\label{tab:idxgen}
\end{table}


\begin{table}
\centering
\begin{tabular}{l|r|r|r|r}
	index type               & CO & IL & MD & WA \\ \hline
	geography $8 \times 8$   &   <0.01\% &   0.16\% &   3.62\% &     1.11\% \\
	geography $16 \times 16$ &   <0.01\% &   4.66\% &  22.43\% &     3.14\% \\
	HTM level 12             &   0.01\% &   1.71\% &   4.82\% &     1.30\% \\
	HTM level 14             &   <0.01\% &   0.18\% &   1.84\% &     0.47\% \\
	HTM level 16             &   <0.01\% &   0.04\% &   0.53\% &     0.23\%
\end{tabular}
\caption{False positive rate of spatial join pre-filtering for the U.S. states Colorado, Illinois, Maryland and Washington. Note, that false positive rates depend on the actual distribution of tweets and not only on the geometry of the states.}
\label{tab:fp}
\end{table}

\begin{table}
\centering
\begin{tabular}{l|r|r|r|r}
   index type               & 300k [s] & 1M [s] & 5M [s] & 1G [s] \\ \hline
   geography $8 \times 8$   &   223 &   780 &  5009 &   -   \\
	geography $16 \times 16$ &   223 &   883 &  4053 &   -   \\
	HTM level 12             &     1 &     1 &     2 & 194   \\
	HTM level 14             &     2 &     2 &     5 & 266   \\
	HTM level 16             &     7 &     4 &     5 & 232
\end{tabular}
\caption{Pre-filtering time for the spatial join query}
\label{tab:prefilter}
\end{table}

\begin{table}
\centering
\begin{tabular}{l|r|r|r|r}
   index type               & 300k [s] & 1M [s] & 5M [s] & 1G [s] \\ \hline
   geography $8 \times 8$   &   295 &   915 &  4276 &   -   \\
	geography $16 \times 16$ &   301 &   773 &  4273 &   -   \\
	HTM level 12             &    12 &    24 &   139 & 2370  \\
	HTM level 14             &     7 &    13 &    58 & 1299  \\
	HTM level 16             &     8 &    10 &    42 & 1032   
\end{tabular}
\caption{Total time for the spatial join query}
\label{tab:total}
\end{table}

\section{Conclusions}

In this paper, we investigated the feasibility of efficient classification of GPS coordinates of Twitter messages by geographic regions using a relational database management system, Microsoft SQL Server~2012. We evaluated the performance of the built-in spatial indexing technology side by side with a customized solution based on Hierarchical Triangular Mesh (HTM) indexing. The built-in spatial index was found to be inadequate to perform spatial joins between large sets of GPS coordinates (on the scale of billions) and complex geographic regions. We showed that our solution, a heuristic combination of existing techniques for handling spatial data in a relational database environment, can easily be a hundred times faster and makes the computation of the aforementioned spatial join available in reasonable time. We pointed out that the strength of HTM indexing is the great control the database programmer has on both the index structure and query plans (via hints). We also demonstrated that aggressive pruning of spatial indices is not a good idea when indexing of very complex regions is a requirement, as range-join-based pre-filtering is significantly faster than exact containment testing, even in case of millions of index entries. To make exact containment testing even faster, we pre-computed the intersections of the complex geographic regions and partial HTM trixels and use these much smaller shapes to filter out false positives.


\section{Acknowledgments}
	
	The authors thank the partial support of the European Union and the European Social Fund through project FuturICT.hu
	(grant no.: TAMOP-4.2.2.C-11/1/KONV-2012-0013),
	and the OTKA 103244 grants.
	EITKIC 12-1-2012-0001 project was partially supported by the Hungarian Government,
	managed by the National Development Agency, and financed by the Research and
	Technology Innovation Fund and the MAKOG Foundation.


\end{document}